\documentclass[12pt, notitlepage]{article}
\pdfoutput=1

\usepackage[toc,page]{appendix}

\usepackage{float}
\usepackage{subfig}
\usepackage{graphicx}
\usepackage[T1]{fontenc}
\usepackage[utf8]{inputenc}
\usepackage{authblk}
\usepackage{amsmath,bm}
\usepackage[left=3cm,right=3cm,top=2cm,bottom=2cm]{geometry}
\usepackage{verbatim}

\renewcommand\footnotemark{}
\begin{document}

\title{Searching for a continuum limit in causal dynamical triangulation quantum gravity}         

\author[a,b]{J.~Ambjorn}
\author[a,c]{D.~Coumbe}
\author[c]{J.~Gizbert-Studnicki}
\author[c]{J.~Jurkiewicz}
\affil[a]{\small{The Niels Bohr Institute, Copenhagen University, \authorcr Blegdamsvej 17, DK-2100 Copenhagen Ø, Denmark. \authorcr E-mail: ambjorn@nbi.dk, daniel.coumbe@nbi.ku.dk.\vspace{+2ex}}} 

\affil[b]{\small{IMAPP, Radboud University, \authorcr Nijmegen, PO Box 9010, The Netherlands.\vspace{+2ex}}}

\affil[c]{\small{Institute of Physics, Jagiellonian University, ul. prof. Stanislawa Lojasiewicza 11, PL 30-348 Krakow , Poland. \authorcr Email: jakub.gizbert-studnicki@uj.edu.pl, jerzy.jurkiewicz@uj.edu.pl.}}


\date{\small({Dated: \today})}          
\maketitle


\begin{abstract}

We search for a continuum limit in the causal dynamical triangulation (CDT) approach to quantum gravity by determining the change in lattice spacing using two independent methods. The two methods yield similar results that may indicate how to tune the relevant couplings in the theory in order to take a continuum limit. 


\vspace{1cm}

\end{abstract}


\begin{section}{Introduction}\label{intro}

Causal dynamical triangulations (CDT) define a nonperturbative approach to quantum gravity in which spacetime is divided into a lattice of 4-dimensional Lorentzian triangles. At first sight this seems to suggest that CDT defines a discrete spacetime, however the hope is that one can study spacetime on lattices of ever decreasing edge length, with the eventual aim of investigating the properties of spacetime in the continuum limit. It is therefore important to be able to reliably determine the lattice spacing in CDT, and how it changes within the parameter space. In this work we calculate the lattice spacing using two independent methods at a number of different points in the parameter space with the aim of gaining an insight into how one might take a continuum limit in CDT. 



As is well known, gravity as a perturbative quantum field theory is not renormalizable by power-counting~\cite{Goroff:1985th}. However, as suggested in Weinberg's seminal work \cite{Weinberg79} the definition of renormalizability might be generalised to the nonperturbative regime as detailed by the asymptotic safety scenario~\cite{Reuter:2012id}. The asymptotic safety scenario would be realised if a finite number of couplings terminated at an ultra-violet fixed point (UVFP), so that the theory remains finite and predictive even in the infinite energy limit, and there is by now a growing body of evidence supporting the existence of such an UVFP~\cite{Lauscher:2001ya,Litim:2003vp,Codello:2007bd,Codello:2008vh,Benedetti:2009rx}. In this way a lattice theory of gravity, such as CDT, might also be used to provide evidence for asymptotic safety by searching for a second-order critical point in its parameter space which would correspond to an UVFP. The divergent correlation length characteristic of a second-order critical point would allow for the possibility of taking the lattice spacing to zero while simultaneously keeping observable quantities fixed in physical units.    

A defining feature of CDT is the existence of a causality condition that allows one to distinguish between space-like and time-like links on the lattice. The existence of a causality condition allows the foliation of the lattice into space-like hypersurfaces of fixed topology. There exist two types of 4-dimensional triangulations in CDT, the $(4,1)$-simplex and the $(3,2)$-simplex, where the notation $(i,j)$ refers to the number of vertices $i$ on hypersurface $t$, and the number of vertices $j$ on hypersurface $t\pm1$. In CDT the lattice spacing of time-like links $a_{t}$ and space-like links $a_{s}$ are not necessarily equal, but are related via an $\alpha$ parameter

\begin{equation}\label{Eqalpha}
a_{t}^{2}=-\alpha a_{s}^{2}.
\end{equation}

\noindent For $\alpha=-1$ we have $a_{t}=a_{s}$. By taking $\alpha\neq -1$ and following Regge's method for describing piecewise linear geometries \cite{Regge:1961px} we obtain a simple expression for the Einstein-Hilbert action of CDT \cite{Ambjorn:2001cv},
\begin{equation}\label{SRegge}
S_{E}=-\left(\kappa_{0}+6\Delta\right)N_{0}+\kappa_{4}\left(N_{(4,1)}+N_{(3,2)}\right)+\Delta \ N_{(4,1)} \  ,
\end{equation} 

\noindent where $\kappa_{0}$ is proportional to $a_{s}^{2}/ G$ and $G$ is the bare Newton's constant. The parameter $\Delta$ is related to the ratio of space-like and time-like links on the lattice, and $\kappa_{4}$ is related to the cosmological constant, which is fixed so one can extrapolate to the infinite volume limit. We can then independently vary $\kappa_{0}$ and $\Delta$ in order to explore the parameter space of CDT.


The key features of the CDT parameter space have by now been largely mapped out. It has been demonstrated that phase C has semiclassical features that closely resemble 4-dimensional de Sitter space \cite{Ambjorn:2007jv,Ambjorn:2008wc}. The solid red line in Fig. \ref{PDcdt} originally thought to separate phase B from phase C appears to be a second-order transition \cite{Ambjorn:2011cg}, allowing for the possibility of taking a continuum limit. However, the location of the newly discovered bifurcation phase \cite{Ambjorn:2014mra} separating phase C from phase B may prevent the possibility of approaching the second order transition line from within the physically interesting phase C (see Fig. \ref{PDcdt}). Furthermore, recent studies suggest the continuum limit may exist for sufficiently large $\kappa_{0}$ and $\Delta$, rather than near the B-C transition as previously thought \cite{Ambjorn:2014gsa}. The aim of this work is to independently check the findings of Ref. \cite{Ambjorn:2014gsa} by determining how the lattice spacing changes as a function of the bare parameters $\kappa_{0}$ and $\Delta$ using two independent methods.

\end{section}


\begin{section}{Simulation details}\label{details}





The first of the two methods we use to determine the effective lattice spacing in this work is based on analysing fluctuations of three-volume, as first introduced in Refs.~\cite{Ambjorn:2008wc} and~\cite{Ambjorn:2007jv}. This method relies on the observation that within phase C of the CDT parameter space the distribution of three-volume as a function of time has an expectation value that closely matches de Sitter space. Since the classical solution for de Sitter space does not contain Newton's constant but the semiclassical fluctuations about de Sitter space do, one can exploit this fact to estimate the lattice spacing in Planck units for each point in phase C of the parameter space. We defer a detailed discussion of this method to section~\ref{method1}. The second of our two methods involves measurements of the so-called spectral dimension, a measure of the effective fractal dimension of a geometry. In CDT, the spectral dimension $D_{S}\left(\sigma\right)$ is defined via a discrete diffusion process, and is related to the probability $P_{r}\left(\sigma\right)$ that a random walk on the geometry returns to its origin after $\sigma$ diffusion steps. In this work we calculate the spectral dimension $D_{S}\left(\sigma\right)$ as a function of $\sigma$ for a number of different points in phase C of the CDT parameter space. As detailed in section~\ref{method2}, a comparison of $D_{S}\left(\sigma\right)$ at different points in the parameter space can then be used to determine the effective change in lattice spacing.  


We aim to reduce systematic errors associated with our measurements in two ways. First, we aim to reduce finite size effects. In Ref. \cite{Coumbe:2014noa}, finite size effects were shown to be negligible for lattice volumes of $N_{4,1}=160$k for the bare parameters $\left(\kappa_{0}=2.2,\Delta=0.6\right)$, $\left(\kappa_{0}=3.6,\Delta=0.6\right)$, and  $\left(\kappa_{0}=4.4,\Delta=0.6\right)$. However, for bare parameters corresponding to finer lattice spacings finite-size effects appeared to increase. For this reason we simulate with a larger lattice volume $N_{(4,1)}=300$k close to the C-A transition line in order to reduce finite size effects. Secondly, the discretisation effect of odd-even oscillations in the short distance spectral dimension have been removed by omitting values of $D_{S}\left(\sigma\right)$ from the fit-functions wherever they become significant as indicated in Fig. \ref{SpecCAline}, typically this occurs around $\sigma\approx 80$.

We reduce statistical errors by ensuring our configurations are thermalized. This check is made by analysing $D_{S}\left(\infty\right)$ as a function of Monte Carlo time and showing there is no statistical difference between the first and second half of the measured data set, as detailed in Ref. \cite{Coumbe:2014noa}.

The spectral dimension in this work is determined by taking the starting point of diffusion processes to be in the time slice containing the maximal number of (4,1) simplices, so as to ensure that we are probing the bulk geometry with each diffusion. The maximal number of time steps $\sigma$ in our calculations is set to 500, and the time extension of our ensembles is $t=80$. We use a linear volume fixing constraint $\delta S=\epsilon|N_{(4,1)}-N_{(4,1)}^{\rm{target}}|$, with $\epsilon=0.02$ after thermalization when determining the change in lattice spacing described in section~\ref{method2}. For technical convenience we use a quadratic volume fixing constraint $\delta S=\epsilon \left( N_{(4,1)}-N_{(4,1)}^{\rm{target}}\right) ^{2}$, with $\epsilon=0.00001$ when determining the lattice spacing described in section~\ref{method1}.\footnote{In order to determine the absolute lattice spacing we fit the parameters of the effective action as described in section \ref{method1}. This is done by analysing the inverse of the covariance matrix of spatial volume fluctuations. To get rid of the zero mode and make  the matrix invertible we allow for total volume fluctuations around   $N_{(4,1)}^{\rm{target}}$ and subtract the effect of volume fixing from the effective action. For the quadratic volume fixing the effect is a simple  shift of the inverse matrix elements by a constant $2 \epsilon$. }

 
The $\left(\kappa_{0},\Delta\right)$ coordinates of the points at which we simulate in this work are schematically depicted in Fig. \ref{PDcdt}. 

\begin{figure}[H]
\centering
\includegraphics[width=0.6\linewidth,natwidth=610,natheight=642]{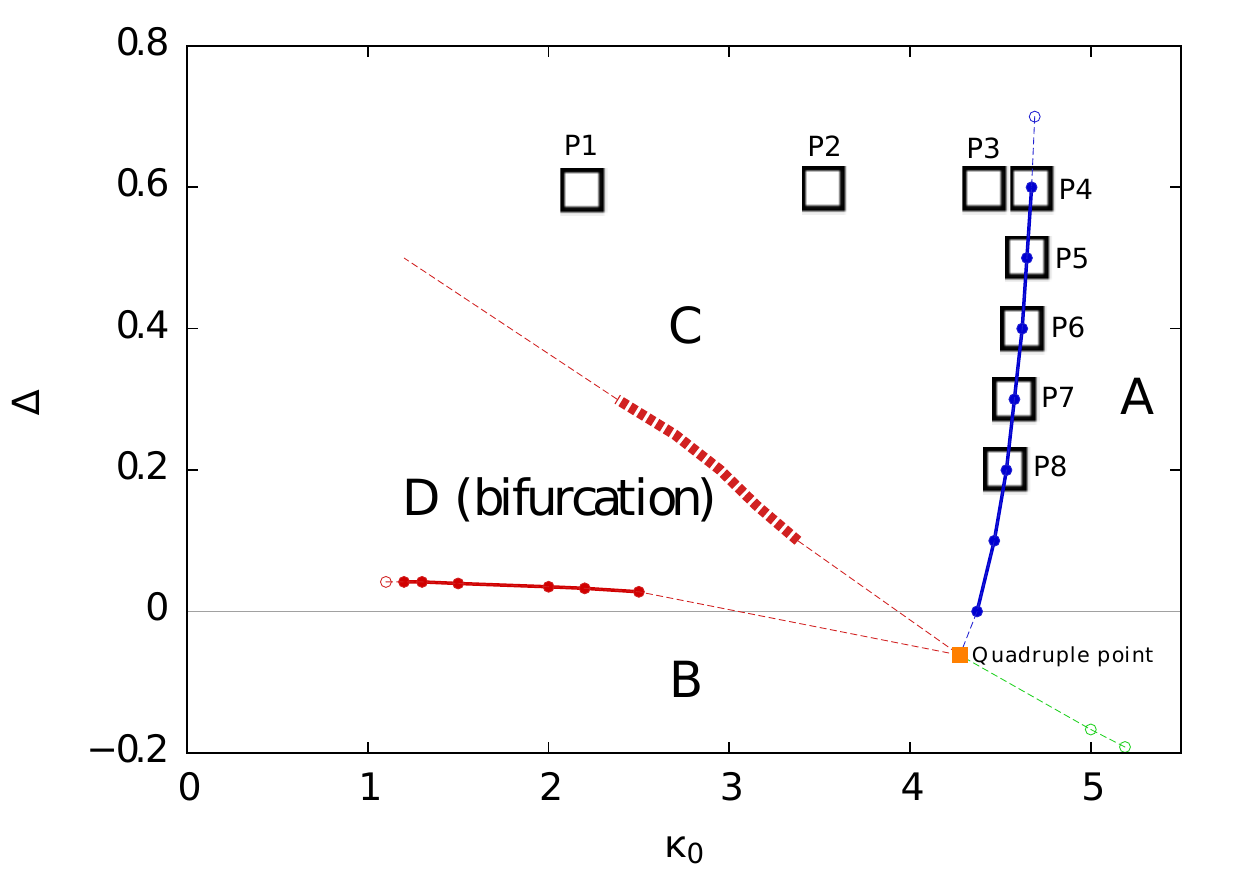}
\caption{\small A schematic representation of the parameter space of CDT. The coloured circles represent measured phase transition points and the coloured lines their interpolation. The open black squares denote the 8 points in the parameter space which we study in this work.}
\label{PDcdt}
\end{figure}

\end{section}


\begin{section}{Method 1: Fluctuations about de Sitter space}\label{method1}


In order to perform numerical simulations one typically has to work with dimensionless quantities by expressing all parameters in terms of the (absolute) lattice spacing $a_{abs}$ and then assuming $a_{abs}=1$. As a result, in order to translate (dimensionless) numerical results  into (dimensionful) physical units one should find a way to measure the absolute lattice spacing. This can be achieved by looking at effective observables and comparing them to corresponding physical constants  which reintroduces a scale into  the numerical system.

The method for determining the absolute lattice spacing that we shall study in this section was first proposed in Refs. \cite{Ambjorn:2007jv,Ambjorn:2008wc}. It is based on the observation that inside phase C of the CDT parameter space, the distribution of volume as a function of time   has an expectation value that closely matches de Sitter space, and hence closely resembles a maximally symmetric spacetime with a positive cosmological constant with 3-volume distributed according to the universal curve  \cite{Ambjorn:2008wc}
\begin{equation}\label{N3t}
\langle N_{3}\left(t\right)\rangle=\frac{3}{4}\frac{N_{(4,1)}}{ s_0 {N_{(4,1)}}^{1/4}}\text{cos}^{3}\left(\frac{t}{ s_0 N_{(4,1)}^{1/4}}\right) .
\end{equation}

$N_{3}(t)$ is (twice) the number of tetrahedra\footnote{Note that in order to have  $\sum_t N_3(t) = N_{(4,1)}$ we set  $N_3(t)\equiv N_{(4,1)}(t) = 2 N_{tetrahedra}(t)$.}
 comprising each spatial slice at a (discrete) time $t$, where the result is
independent of the total lattice volume. The constant $s_0$ depends on the radius of the extended part of the universe (the so-called {\it{blob}}).
At the same time  quantum fluctuations $\delta N_3(t) = N_3(t) -  \langle N_{3}\left(t\right) \rangle $  are consistent with the effective action \cite{Ambjorn:2008wc}
\begin{equation}\label{Seff}
 S_{eff}= \frac{1}{\Gamma}\sum_t \left(\frac{\Big( N_3(t+1)-N_3(t)\Big)^2}{N_3(t+1)+N_3(t)} + \mu N_3(t)^{1/3} - \lambda N_3(t) \right) ,
\end{equation}
where $\Gamma$ is a dimensionless constant depending on the amplitude of quantum fluctuations.

In this section we assume the CDT universe measured inside phase C is that of Euclidean de Sitter space (four-sphere) with superimposed quantum fluctuations of the spatial volume obtained for a spatially isotropic and homogeneous  metric, having a line element

\begin{equation}\label{ds2}
ds^2 = g_{\tau\tau} d\tau^2 + a^2(\tau) d\Omega^3 .
\end{equation}

The Einstein-Hilbert action calculated for the metric (\ref{ds2})  becomes a minisuperspace (MS) action which can be parametrised by a  spatial volume observable 
$V_3(\tau)=\int d\Omega_3\sqrt{g|_{S^3}}=2\pi^2 a^3(\tau)$ in the following form
\begin{equation}\label{CSMSV}
S_{MS}=\frac{1}{24 \pi G}\int d\tau  \sqrt{g_{\tau\tau}}  \left(  \frac{{g^{\tau\tau}} \left( { \partial_\tau V_3(\tau)}  \right)^2}{V_3(\tau)}+ \tilde \mu V_3(\tau)^{1/3}-\tilde \lambda V_3(\tau) \right)  ,
\end{equation}
where $G$ is the dimensionful ($[G] = L^2$) Newton's constant. For the MS action (\ref{CSMSV})  the semiclassical  spatial volume profile is given by

\begin{equation}\label{V3t}
\langle  V_{3}\left(\tau\right)\rangle= 2 \pi^2 { \cal R}^3 \text{cos}^{3}\left(\frac{\sqrt{g_{\tau\tau}}\ \tau}{ {\cal R}}\right)  =\frac{3}{4}\frac{V_4}{\tilde s_0 V_4^{1/4}}\text{cos}^{3}\left(\frac{ \sqrt{g_{\tau\tau}}\ \tau}{\tilde s_0 V_4^{1/4}}\right) ,
\end{equation}
where $ {\cal R}$ is the radius and $V_4 = \frac{8 \pi^2}{3} {\cal R}^4$ is the total 4-volume of the four-sphere (and $ \tilde s_0 = \left(\frac{3}{8 \pi^2}\right)^{1/4}$).

A natural identification of discrete  expressions  (\ref{N3t}) and (\ref{Seff})   with their continuous counterparts (\ref{V3t}) and (\ref{CSMSV}), respectively, via simple dimensional analysis leads to the conclusion that
\begin{equation}\label{lpl2}
 l_{pl}^2 \equiv G\propto \Gamma \cdot a_{abs}^2 ,
\end{equation}
where $l_{pl}=\sqrt{G}$ is the Planck length (in units $\hbar=c=1$) and  $a_{abs}$ is the physical lattice spacing. The dimensionless proportionality factor in (\ref{lpl2}) is derived in Appendix 1, giving the  formula for  absolute lattice  spacing that we will use in this section, namely
\begin{equation}\label{aabs}
a_{abs}=  \sqrt{\frac {{3 \sqrt{6}}} {\sqrt{C_4} \ s_0^2 \  \Gamma }  } l_{pl} ,
\end{equation}
where $C_4$ is the dimensionless {\it effective} volume of a unit  4-simplex, 
while $s_0$ is the extension of the universe defined by Eq. (\ref{N3t}) and $\Gamma$ is the amplitude of quantum fluctuations defined by Eq. (\ref{Seff}).  In order to  estimate $s_0$ we make a fit of Eq. (\ref{N3t}) to the average volume profile $\langle N_3(t) \rangle$ measured in numerical simulations. The procedure of estimating   $\Gamma$  is based on the analysis of the covariance matrix $C_{t\,t'} \equiv \langle\delta N_3(t) \delta N_3(t') \rangle$, where $\delta N_3(t) \equiv N_3(t)-\langle N_3(t) \rangle$, and  follows the procedure described in detail in Ref. \cite{Ambjorn:2011ph}. We use the inverse of the  empirical  covariance matrix $C^{-1}_{t\,t'}$, whose elements are  (in a semiclassical approximation) given by second order derivatives of the effective action (\ref{Seff}), to fit  the action parameters to  numerical data.

We use three  methods for determining $a_{abs}$. Each method uses formula (\ref{aabs}) but differs in the way $C_4$ is determined (the measurement of $s_0$ and $\Gamma$ is universal). The three methods are as follows (see Appendix 1 for more details):
\begin{itemize}
\item{{\bf In method (a)} we set $C_4 = \text{const.}$  (independent of the position in  CDT parameter space). This is consistent with the assumption that  spatial layers of equal (integer) $t$, which are built of equilateral tetrahedra, are separated by a universal time distance of constant lattice length. As such, tetrahedra are constituents  (faces) of (4,1)-simplices, in this approach one takes into account only simplices of this  type, disregarding all (3,2)-simplices.}
\item{{\bf In method (b)} we assume: $C_4 = C_4(\xi)$, where: $\xi = N^{(3,2)}/ N^{(4,1)}$. In Ref. \cite{Ambjorn:2011ph} it was shown that (3,2) simplices form closed layers which can be attributed a half-integer $t$ variable. The number of (3,2) simplices in each layer is consistent with the volume profile of (4,1)-simplices (tetrahedra) in integer $t$ by  a simple rescaling of $\xi$. This method is more general than (a), as it takes into account both  (4,1) and (3,2) simplicial layers. A simplification and approximation comes from the fact that one assumes that all 4-simplices are symmetric and thus the volume of each 4-simplex is constant and universal, independent of the position in the parameter space ($C_{41}=C_{32}=\text{const.}$).}
\item{{\bf In method (c)} we assume: $C_4 = C_4(\xi, C_{41},C_{32} ) $, where: $\xi = N^{(3,2)}/ N^{(4,1)}$. This is the most general assumption  as it accounts for the fact that the volume of a unit 4-simplex depends on the position in the bare parameter space of CDT and, in principle, $ C_{41}(\alpha) \neq C_{32}(\alpha) $. This method requires determining the value of the parameter  $\alpha=\alpha(\kappa_0,\Delta,\kappa_4)$ which defines the asymmetry between the length of time-like and space-like links on the lattice.
}
\end{itemize}

The reason for using various methods is due to the fact that  we want to compare the results of the absolute lattice spacing derived in this section with the relative lattice spacing defined by the rescaling of the spectral dimension (see Section \ref{method2}), and check which method gives the closest agreement. Additionally, method (c), which naively seems to be the most accurate, breaks down close to the C-A phase transition (where double valued or complex $\alpha$ solutions are possible).   

In principle, all three parameters in formula (\ref{aabs}): $s_0$, $\Gamma$ and $C_4$ (in methods (b) and (c)) depend  on the actual position in the CDT bare parameter space. As a result $a_{abs} = a_{abs}(\kappa_0,\Delta,\kappa_4)$. One can check how $a_{abs}$ changes by following various trajectories in the  parameter space and determining those for which $a_{abs}\to 0$ monotonically. The values of $a_{abs}$ at each sampled point of the bare parameter space in phase C are presented in Table \ref{TableComparison0}  and Fig. \ref{FigAbs},  where we compare the absolute lattice spacing  calculated using the three different methods described. The error bars shown in Table \ref{TableComparison0} and Fig. \ref{FigAbs} are related to fitting errors of parameters $\Gamma$, $s_0$ and $\xi$ used to calculate $a_{abs}$ by formula (\ref{aabs}) and do not take into account statistical errors. The error bars clearly increase close to the C-A phase transition  which is mainly  due to the increased error associated with $\Gamma$. $\Gamma\to\infty$ as one approaches  the phase transition resulting in a high noise/signal ratio in the measured inverse covariance matrix, which produces the increased error estimate.  

\begin{table}                                                                                                                                                                 
\begin{center}                                                                                                                                                                
\begin{tabular} {|c|c|c|c|c|}                                                                                                                                                 \hline                                                                                                                                                                        
{Label in Fig.~\ref{PDcdt}} & {${(\kappa_0,\Delta)}$} & {$a_{abs}$ \bf(a)}   & {$a_{abs}$ \bf(b)} &{$a_{abs}$ \bf(c)}\\ \hline
\hline                                                                                                                                                                        
P1&$(2.20,0.6)$&$ 1.82\pm0.01$&$1.48\pm0.01$ &$2.01\pm0.01$ \\ \hline
P2&$(3.60,0.6)$&$ 1.34\pm0.01$&$1.18\pm0.01$ &$1.52\pm0.01$ \\ \hline
P3&$(4.40,0.6)$&$ 0.98\pm0.02$&$0.92\pm0.02$ &$1.12\pm0.02$ \\ \hline
P4&$(4.67,0.6)$&$ 0.56\pm0.06$&$0.55\pm0.06$ &$0.65\pm0.06$ \\ \hline
P5&$(4.64,0.5)$&$ 0.61\pm0.05$&$0.60\pm0.05$ &$0.70\pm0.05$ \\ \hline
P6&$(4.62,0.4)$&$ 0.60\pm0.04$&$0.58\pm0.04$ &$\ \ 0.68\pm0.04\, ^*$ \\ \hline
P7&$(4.57,0.3)$&$         0.63\pm0.04$&$0.61\pm0.04$ &$\ \ 0.71\pm0.04\, ^*$ \\ \hline
P8&$(4.53,0.2)$&$ 0.65\pm0.03$&$0.64\pm0.03$ &$\ \ 0.73\pm0.03\, ^*$ \\ \hline
\end{tabular}                                                                                                                                                                 
\end{center}                                                                                                                                                                  
\caption{\small                                                                                                                                                              
A table showing $a_{abs}$  for 8  points $(\kappa_0,\Delta)$ in phase C of the  CDT parameter space.  The absolute lattice spacing $a_{abs}$ was determined by  analysing fluctuations about de Sitter space using three different methods (a), (b) and (c)  described in the text. The lattice spacing is in units of the Planck length $l_{pl}$. \newline
$^*$ For the last three points the method (c) breaks down, as there exist two real solutions for $\alpha$ (with $\alpha<1$ and $\alpha\gg 1$) and thus  two solutions for $a_{abs}$ (among the two we have chosen the one with $\alpha<1$, which is also the case  for the points where only one solution is possible - this fact is denoted by: $^*$ ).
}
\label{TableComparison0}                                                                                                                                                     
\end{table}

\begin{figure}[H]
\centering
\includegraphics[width=0.6\linewidth,natwidth=610,natheight=642]{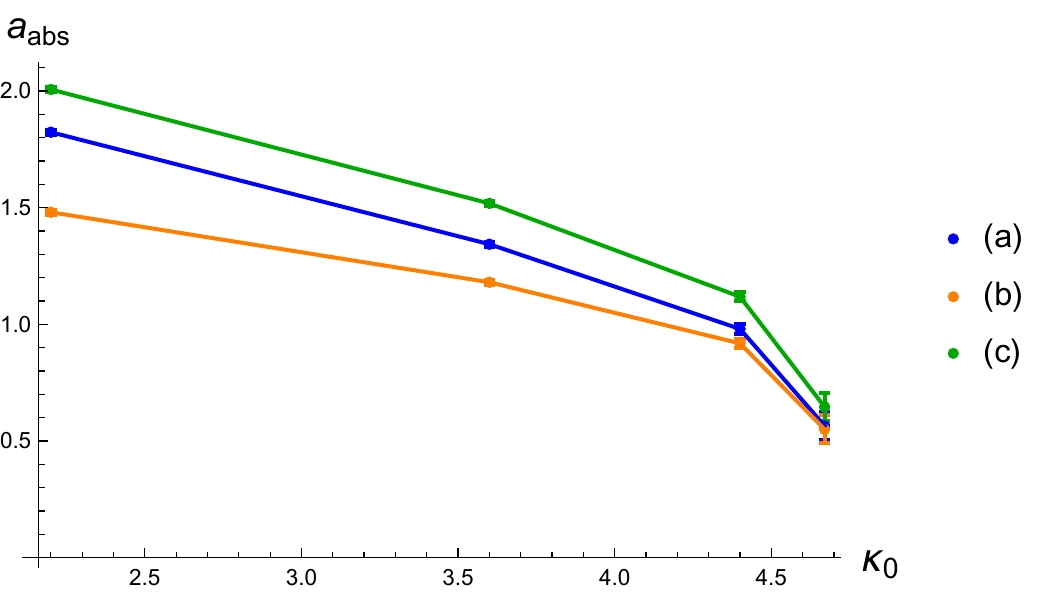}
\includegraphics[width=0.6\linewidth,natwidth=610,natheight=642]{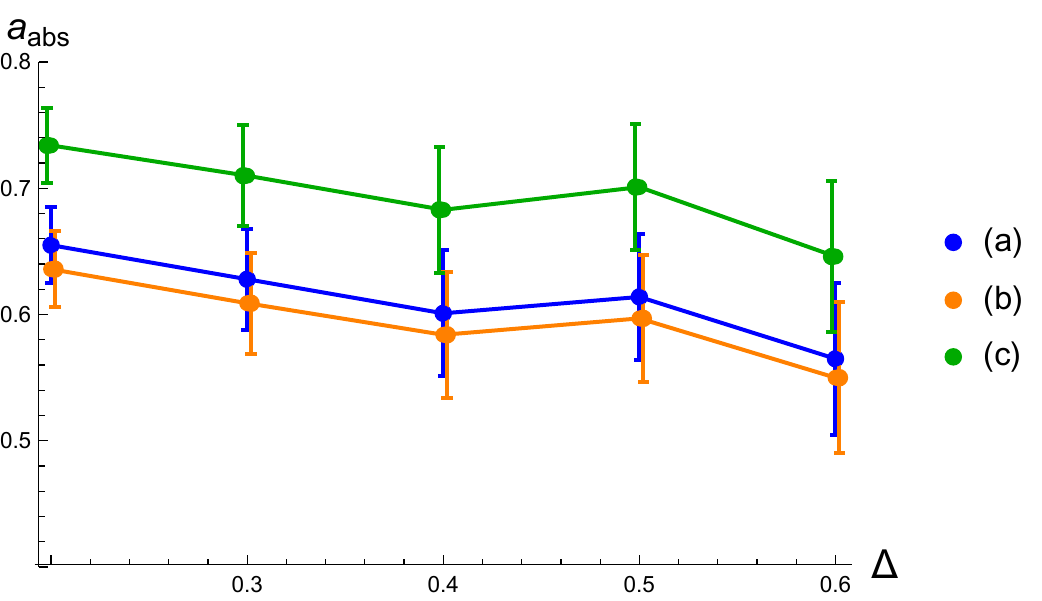}
\caption{\small Visualisation of the data in Table \ref{TableComparison0}. The top chart shows the dependence of $a_{abs}$ on $\kappa_0$ for fixed $\Delta=0.6$, and the bottom chart shows the dependence of $a_{abs}$ on $\Delta$ along the C-A phase transition line. Note that the scale is different in each chart.}
\label{FigAbs}
\end{figure}


\end{section}


\begin{section}{Method 2: Rescaling of the spectral dimension}\label{method2}




The spectral dimension is related to the probability $P_{r}\left(\tilde\sigma\right)$  that a random walk returns to the origin after a fictitious diffusion time  $\tilde \sigma$. The spectral dimension can be derived from the $d$-dimensional diffusion equation

\begin{equation}
\frac{\partial}{\partial\tilde\sigma}K_{g}\left(\zeta_{0},\zeta,\tilde\sigma\right)-g^{\mu\nu}\bigtriangledown_{\mu}\bigtriangledown_{\nu}K_{g}\left(\zeta_{0},\zeta,\tilde\sigma\right)=0,
\label{diffusion1}
\end{equation}

\noindent where $K_{g}$ is the heat kernel defining the probability density of diffusing between $\zeta_{0}$ and $\zeta$. $\bigtriangledown$ is the covariant derivative of the metric $g_{\mu\nu}$. For infinitely flat Euclidean space, Eq. (\ref{diffusion1}) has the solution

\begin{equation}
K_{g}\left(\zeta_{0},\zeta,\tilde\sigma\right)=\frac   {\rm{exp}\left(-d_{g}^{2}\left(\zeta,\zeta_{0}\right)/4\tilde\sigma\right)}  {\left(4\pi\tilde\sigma\right)^{d/2}},
\end{equation}

\noindent where $d_{g}^{2}\left(\zeta,\zeta_{0}\right)$ defines the geodesic distance between $\zeta$ and $\zeta_{0}$.

The quantity that is measured in the numerical simulations is the probability of return $P_{r}\left(\tilde\sigma\right)$, which in asymptotically flat space is given by

\begin{equation}
P_{r}\left( \tilde \sigma\right)=\frac{1} {\tilde\sigma^{d/2}}.
\end{equation}

\noindent The scale dependent spectral dimension $D_{S}\left(\tilde \sigma \right)$ is then computed by taking the logarithmic derivative with respect to $\tilde \sigma$, 

\begin{equation}
D_{S}\left(\tilde \sigma \right)=-2\frac{d\rm{log}\langle P_{r}\left(\tilde \sigma\right)\rangle}{d\rm{log}\tilde \sigma}.
\label{SpecDef}
\end{equation}

In CDT, Euclidean space is discretized by a simplicial manifold, and the diffusion time $\tilde\sigma$ is counted by the number of discrete diffusion steps $\sigma$ on the dual lattice, i.e. each diffusion step consists of moving from the centre of one four-simplex to the centre of a neighbouring four-simplex. One starts the diffusion process from a randomly chosen simplex and numerically calculates the probability of return   to the origin $P_{r}\left( \sigma\right)$ after $\sigma$ diffusion steps.  Measured values of $P_{r}\left(\sigma \right)$ are  averaged over different starting points and different triangulations, and then one uses a discrete version of Eq. (\ref{SpecDef}) to calculate $D_{S}\left(\sigma \right)$.


In a lattice formulation of an asymptotically safe field theory an ultraviolet fixed point is expected to manifest as a second-order critical point. At a second-order critical point macroscopic observables become independent of the microscopic regularisation, and should therefore become scale invariant. A scale invariant spectral dimension would appear as a perfectly flat $D_{S}\left(\sigma\right)$ curve over all distance scales $\sigma$. Therefore, we should expect to see progressively flatter spectral dimension curves as we approach such a fixed point. The amount by which we must rescale $D_{S}\left(\sigma\right)$ at a particular point in the parameter space such that it agrees with $D_{S}\left(\sigma\right)$ at another point in the parameter space will then be related to the change in relative lattice spacing when transforming between the two points, a method first proposed in Ref. \cite{Coumbe:2014noa}. 

To test the validity of this method we calculate the spectral dimension for 8 different points in the parameter space (see Fig. \ref{PDcdt}) and compare the results with those obtained using the independent method of calculating the absolute lattice spacing via fluctuations about de Sitter space as described in Section \ref{method1}. Our results for the spectral dimension as a function of diffusion time for the 8 points sampled in the parameter space are presented in Fig. \ref{SpecCAline}. Typically, we find a small scale spectral dimension that is more consistent with the lower bound $D_{S}\left(\sigma \rightarrow 0 \right)\sim 3/2$ (see Ref.~\cite{Coumbe:2014noa} for typical error estimates) than with $D_{S}\left(\sigma \rightarrow 0 \right)\simeq 2$, thereby supporting the findings of Ref. \cite{Coumbe:2014noa}. Furthermore, except for the points close to the C-A transition we find a large scale spectral dimension that is consistent with $D_{S}\left(\sigma \rightarrow \infty\right) \sim 4$.

Within phase C of CDT, the fit function 

\begin{equation}
D_{S}\left(\sigma\right)=a-\frac{b}{c+\sigma}
\end{equation}

\noindent has been shown to accurately fit the spectral dimension data \cite{Ambjorn:2005db,Coumbe:2014noa}, a result that is also supported by purely analytic models \cite{Giasemidis:2012qk}. In Fig. \ref{SpecRescale} we rescale the fit function according to \footnote{In order to compare the results of $a_{rel}$ with the absolute lattice spacing $a_{abs}$ one should rescale $\sigma$ by $a_{rel}^2$ due to the squared covariant derivative of the metric in Equation~(\ref{diffusion1}).} 
\begin{equation}
D_{S}\left(\sigma\right)=a-\frac{b}{c+\sigma /a_{rel}^2},
\end{equation}

\noindent where $a_{rel}$ is chosen such that the curves give the best overlap. The curves are normalized such that the scale factor $a_{rel}$ is set to unity for the point $(\kappa_{0}=2.2, \Delta=0.6)$ (this choice is arbitrary since any pair of $(\kappa_{0},\Delta)$ values could be defined as the point relative to which all other points are compared). The resulting values of $a_{rel}$ and their associated error estimates at each sampled point in the parameter space are shown in Table \ref{TableComparison3}.



\begin{figure}[H]
\centering
\includegraphics[width=0.75\linewidth,natwidth=610,natheight=642]{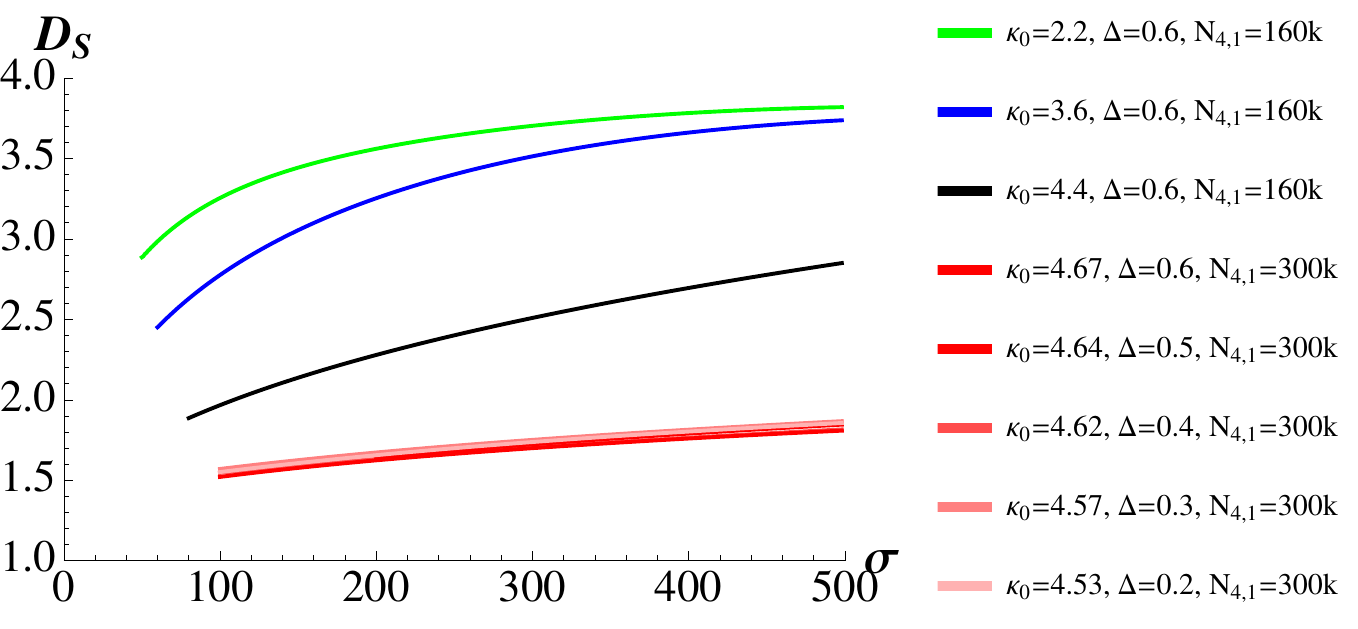}
\caption{\small The spectral dimension $D_{S}\left(\sigma\right)$ as a function of diffusion time $\sigma$ for 8 different points in phase C of the CDT parameter space.}
\label{SpecCAline}
\end{figure}

\begin{figure}[H]
\centering
\includegraphics[width=0.7\linewidth,natwidth=610,natheight=642]{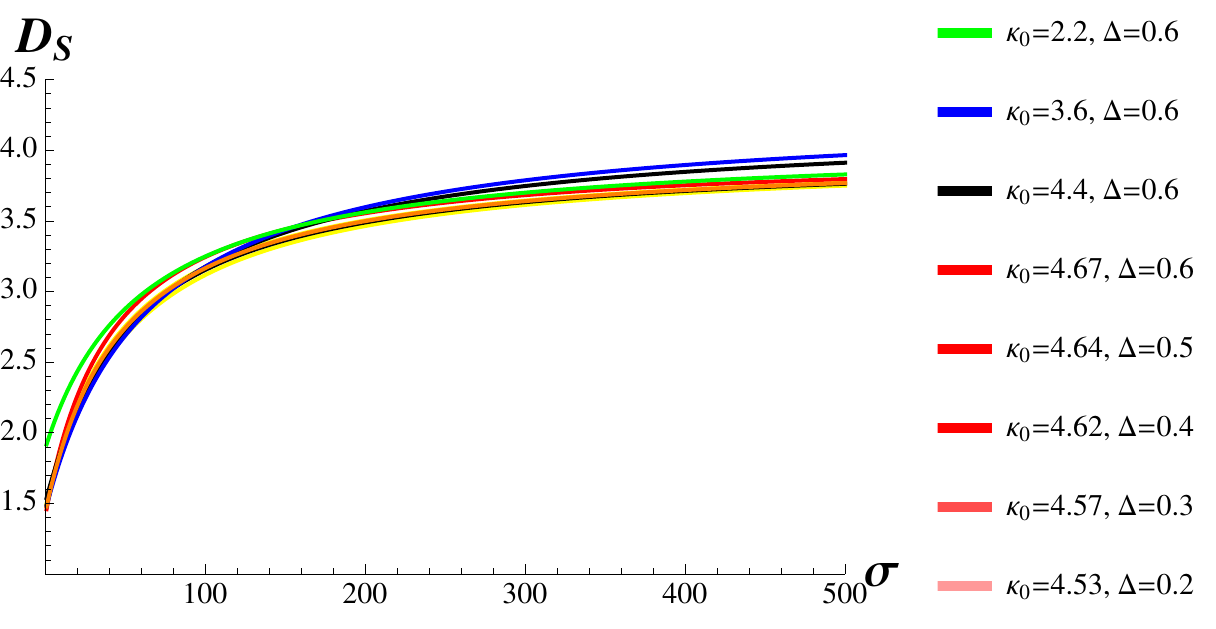}
\caption{\small The spectral dimension fits rescaled according to $D_{S}\left(\sigma\right)=a-\frac{b}{c+\sigma /a_{rel}^{2}}$, where $a_{rel}$ is chosen such that the curves give the best overlap.}
\label{SpecRescale}
\end{figure}




Since we wish to minimise systematic errors associated with determining $a_{rel}$ we make a statistical comparison between the spectral dimension curves we wish to compare. We calculate the standard deviation by comparing central data points for $D_{S}\left(\sigma\right)$ at 10 different $\sigma$ values for our canonical point $(\kappa_{0}=2.2, \Delta=0.6)$ with the rescaled $D_{S}\left(\sigma\right)$ at our comparison point. To reduce discretisation effects we only compare values for which $\sigma \geq 100$. We compare 10 evenly spaced $D_{S}\left(\sigma\right)$ values between $\sigma=100$ and $\sigma=460$. We determine the corrected sample standard deviation $S$ via 

\begin{equation}
S=\sqrt{\frac{\sum\limits_{j=0}^{j=9} \left(D_{S}\left(\sigma=100+40j\right)-D_{S}^{C}\left(\sigma=100+40j\right)\right)^{2}}{n-1}},
\end{equation}

\noindent where $D_{S}^{C}(\sigma)$ is the spectral dimension at the canonical point $(\kappa_{0}=2.2, \Delta=0.6)$ and $D_{S}(\sigma)$ is the rescaled spectral dimension curve with which we make the comparison, e.g. $(\kappa_{0}=3.6,\Delta=0.6)$, and $n$ is the number of compared points. Using this method we can determine how the standard deviation $S$ varies as a function of $a_{rel}$, as shown by the black curve in Fig~\ref{ErrorMethod2}. The value of $a_{rel}$ for which $S$ is minimised, as indicated by the black dashed vertical line in Fig~\ref{ErrorMethod2}, then corresponds to the value of $a_{rel}$ for which the curves give the best overlap. 

It is important to estimate the errors associated with determining the relative lattice spacing. To estimate this error we calculate $S$ as a function of $a_{rel}$ by comparing the minimum possible values of $D_{S}\left(\sigma\right)$ allowed by the error bars with the maximum possible values of $D_{S}^{C}\left(\sigma\right)$ allowed by the error bars, as shown by the red curve in Fig.~\ref{ErrorMethod2}, which is calculated by comparing $D_{S}\left(\sigma\right)$ values at the points $(\kappa_{0}=3.6,\Delta=0.6)$ and $(\kappa_{0}=2.2,\Delta=0.6)$. The difference in the values of $a_{rel}$ for which the red and black curves are minimised then gives an estimate of the error associated with $a_{rel}$. For the points close to the C-A transition we typically find $D_{S}\left(\sigma \rightarrow \infty\right)<4$, and so to aid a better comparison with $D_{S}^{C}\left(\sigma\right)$ we constrain $a \approx 4$ in the fit function while still ensuring a good fit to the data. Furthermore, for the points close to the C-A transition we compare only the central values of $D_{S}\left(\sigma\right)$ with the central and maximal values of $D_{S}^{C}\left(\sigma\right)$, consequently the error estimate associated with $a_{rel}$ for the points close to the C-A transition is likely to be significantly underestimated. 

\begin{figure}[H]
\centering
\includegraphics[width=0.6\linewidth,natwidth=610,natheight=642]{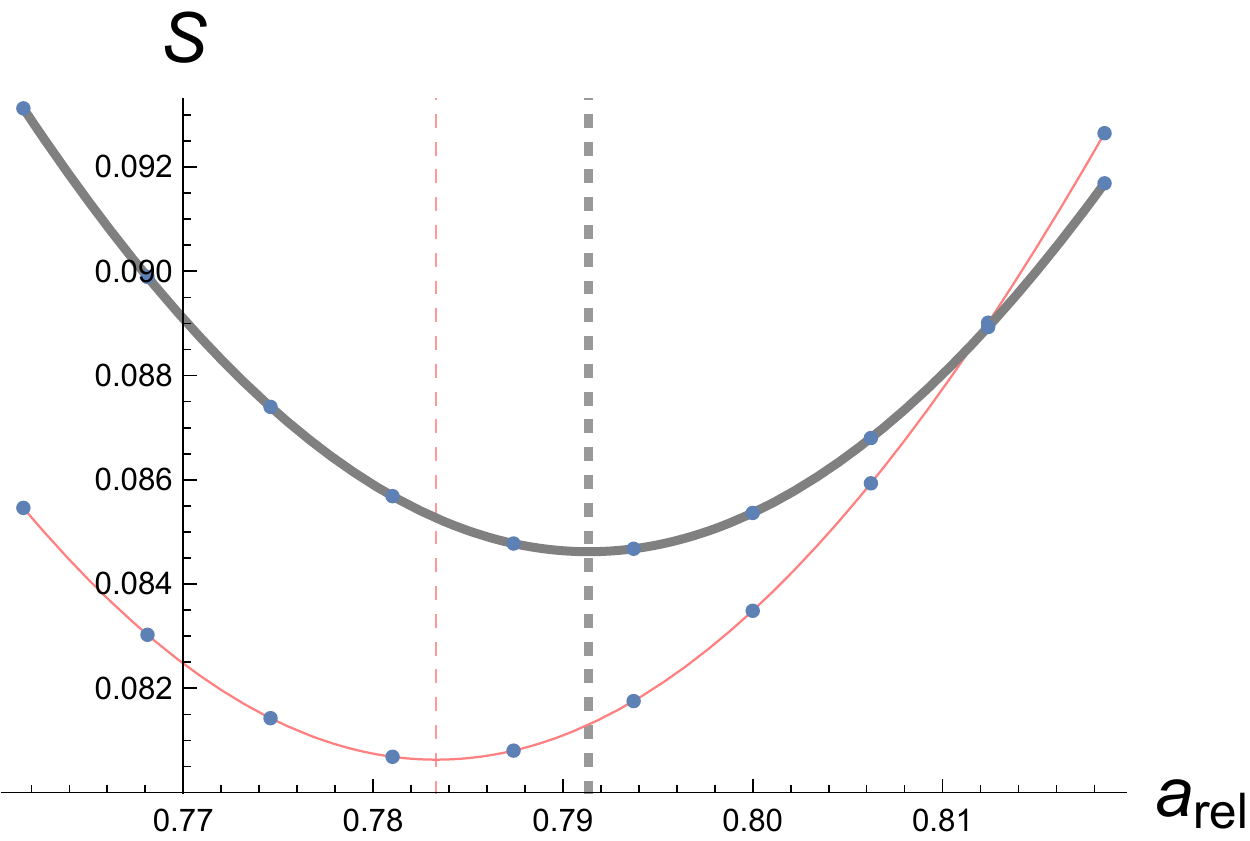}
\caption{\small The standard deviation $S$ (filled dots) calculated by comparing $D_{S}(\sigma)$ at $(\kappa_{0}=2.2,\Delta=0.6)$ with the rescaled curve at $(\kappa_{0}=3.6,\Delta=0.6)$ as a function of $a_{rel}$. An interpolating function has been used to smoothly interpolate between data points. The value of $a_{rel}$ for which the solid black curve is minimised yields the best overlap between the two $D_{S}(\sigma)$ curves. The solid red curve is determined by comparing the minimum values of $D_{S}(\sigma)$ allowed by the error bars at the point $(\kappa_{0}=3.6,\Delta=0.6)$ with the maximal values of $D_{S}(\sigma)$ allowed within error bars at the point $(\kappa_{0}=2.2,\Delta=0.6)$. The difference in the minimum of these two curves gives an estimate of the error associated with $a_{rel}$.}
\label{ErrorMethod2}
\end{figure}

\end{section}


\begin{section}{Indications of a continuum limit}\label{indications}

The change in lattice spacing determined by analysing fluctuations about de Sitter space and by rescaling the spectral dimension both indicate the lattice spacing is strongly dependent on $\kappa_{0}$, but that it is either independent or only very weakly dependent on $\Delta$. The results summarised in Table \ref{TableComparison3} and Fig. \ref{FigRel} suggest that maximising $\kappa_{0}$ within phase C is predominately responsible for minimising the lattice spacing. Since the C-A transition defines the line of maximal $\kappa_{0}$ values in Phase C, it should also then define the set of points for which the lattice spacing is minimised for any given $\Delta$ value. Extrapolating the measured transition points along the C-A line over the entire $\Delta$ range suggests that $\kappa_{0}$ is maximised within phase C for very large, possibly infinite $\Delta$. If this scenario is correct it suggests one should tune the bare parameters to the C-A transition and move in the direction of increasing $\Delta$ in order to approach the continuum limit.\interfootnotelinepenalty=10000 \footnote{\scriptsize It is unlikely that one can take a continuum limit anywhere on the C-A transition itself since it is almost certainly first-order \cite{Ambjorn:2012ij}, however it may be that the C-A transition terminates at a second-order critical point for some $\Delta$ value, just as it appears to do at the quadrupole point (see Fig. \ref{PDcdt}).}

\begin{table}
\begin{center}
\begin{tabular} {|c|c||c||c|c|c|}
\hline
{Label in Fig.~\ref{PDcdt}} & {${(\kappa_0,\Delta)}$}& $a_{rel}$ & {$a_{abs}$ \bf(a)}	& {$a_{abs}$ \bf(b)} &{$a_{abs}$ \bf(c)}\\ \hline
\hline
P1& $(2.20,0.6)$& $1  $&$	1$	&$1$ 	&$1$ \\ \hline
P2& $(3.60,0.6)$& $0.791\pm0.008  $& $0.74\pm0.01  $&$	0.80\pm0.01 $	&$0.76\pm0.01$  \\ \hline
P3& $(4.40,0.6)$& $0.336\pm0.006  $& $0.54\pm0.01  $&$	0.62\pm0.02 $	&$0.56\pm0.01$  \\ \hline
P4& $(4.67,0.6)$& $0.116\pm0.001  $& $0.31\pm0.03  $&$	0.37\pm0.04 $	&$0.32\pm0.03$  \\ \hline
P5& $(4.64,0.5)$& $0.134\pm0.001  $& $0.34\pm0.03  $&$	0.40\pm0.04 $	&$0.35\pm0.03$  \\ \hline
P6& $(4.62,0.4)$& $0.118\pm0.003  $& $0.33\pm0.02  $&$	0.40\pm0.03 $	&$0.34\pm0.02$  \\ \hline
P7& $(4.57,0.3)$& $0.122\pm0.001  $& $0.34\pm0.02  $&$	0.41\pm0.03 $	&$0.35\pm0.02$  \\ \hline
P8& $(4.53,0.2)$& $0.122\pm0.001  $& $0.36\pm0.02  $&$	0.43\pm0.02 $	&$0.37\pm0.02$  \\ \hline
\end{tabular}
\end{center}
\caption{\small 
A table showing $a_{rel}$ for 8 points in phase C of the CDT parameter space. The values of $a_{rel}$ are normalized such that $a_{rel}=1$ for the point $\left(\kappa_{0}=2.2,\Delta=0.6 \right)$, a choice which is of course arbitrary. For comparison we also show the lattice spacing  $a_{abs}$ rescaled (independently for each of the methods (a), (b) and (c)) such that $a_{abs}=1$ for the point $\left(\kappa_{0}=2.2,\Delta=0.6 \right)$.}
\label{TableComparison3}
\end{table}


\begin{figure}
\centering
\includegraphics[width=0.7\linewidth,natwidth=610,natheight=642]{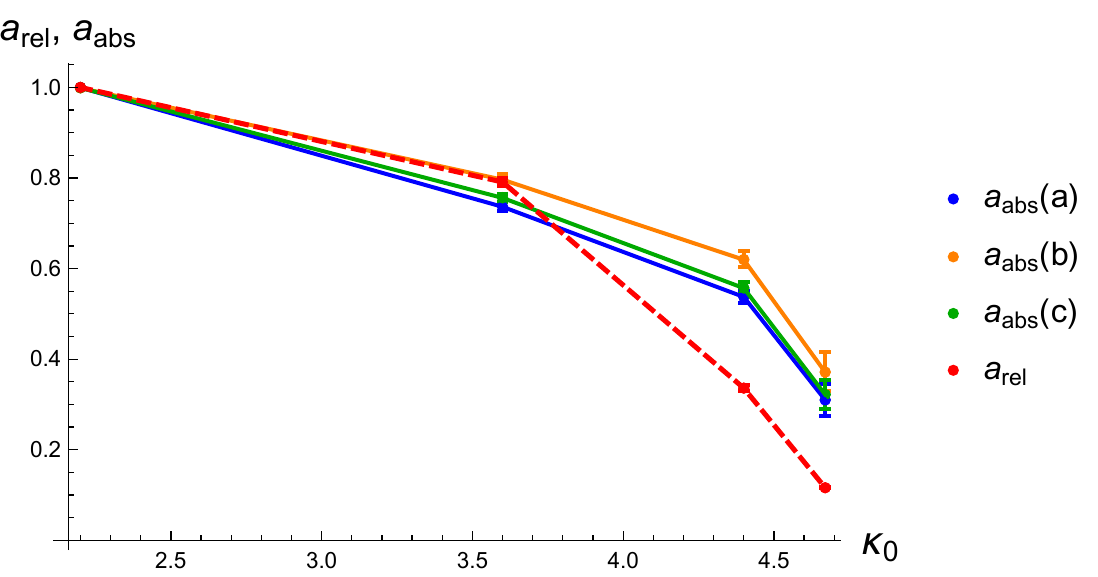}
\includegraphics[width=0.7\linewidth,natwidth=610,natheight=642]{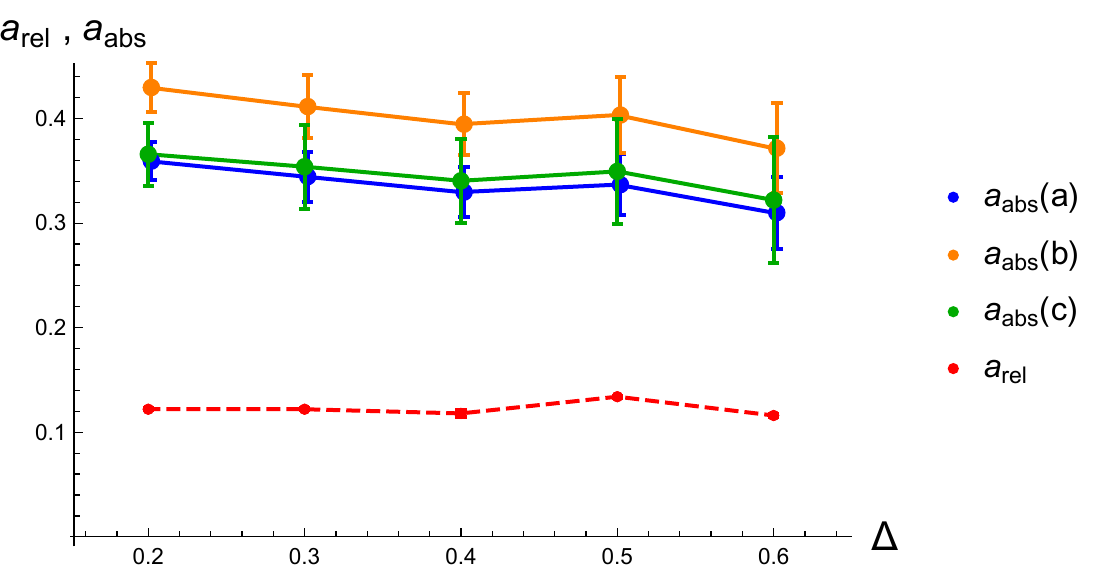}
\caption{\small Visualisation of data from Table \ref{TableComparison3}. The lattice spacing   data are normalized such that $a=1$ for the point $\left(\kappa_{0}=2.2,\Delta=0.6 \right)$. The top chart shows the dependence of $a_{abs}$ and $a_{rel}$ on $\kappa_0$ for fixed $\Delta=0.6$. The bottom chart shows the dependence of $a_{abs}$ and $a_{rel}$ on $\Delta$ close to the C-A phase transition line.}
\label{FigRel}
\end{figure}

The data presented in this work suggests the lattice spacing in phase C is minimised when $\kappa_{0}$ is maximised within phase C, which implies one must tune to the C-A transition line to take a continuum limit. Conversely, the point at which $\kappa_{0}$ is minimised within phase C may then be a candidate for an infra-red fixed point (IRFP). Based on our current picture of the CDT parameter space (Fig. \ref{PDcdt}) such an IRFP would exist on the transition line dividing phase C and the bifurcation phase for the minimal allowed value of $\kappa_{0}$. Our measurements may then be interpreted as suggesting a renormalization group trajectory within the parameter space shown schematically in Fig. \ref{PDcdtRG}. 

\begin{figure}[H]
\centering
\includegraphics[width=0.6\linewidth,natwidth=610,natheight=642]{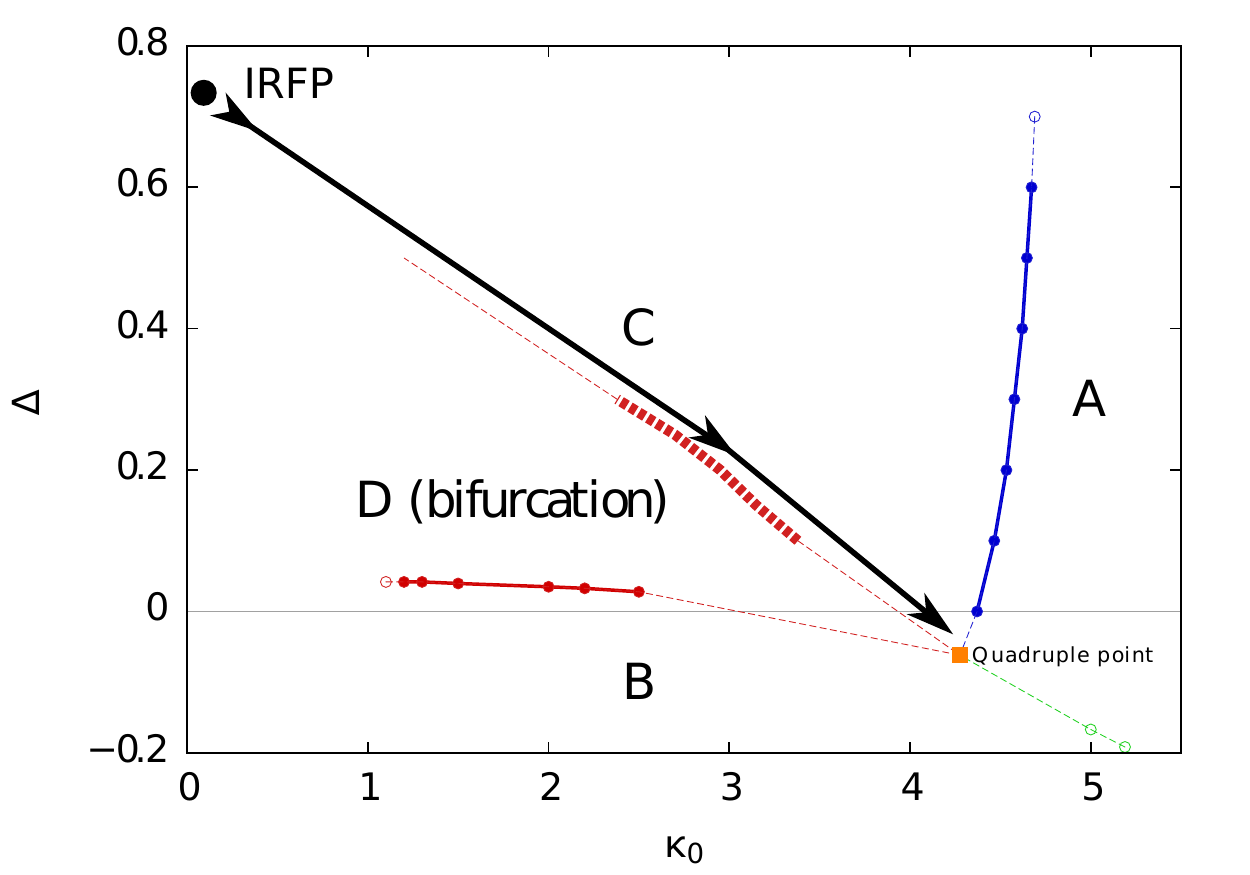}
\caption{\small A possible renormalization group trajectory in phase C of CDT. The trajectory flows from an infra-red fixed point (IRFP) in the direction of decreasing lattice spacing (as indicated by the arrows), as inferred from calculations of $a_{abs}$ and $a_{rel}$ at 8 different points in the parameter space (as shown in Fig.~\ref{PDcdt}).}
\label{PDcdtRG}
\end{figure}


\end{section}


\begin{section}{Discussion and conclusions}\label{conclusions}

If causal dynamical triangulations is to be a viable candidate for a nonperturbative theory of quantum gravity then the expectation is that it should realise the asymptotic safety scenario for gravity. If CDT is to realise the asymptotic safety scenario then it should contain a non-trivial fixed point, which in a lattice formulation such as CDT would appear as a second-order critical point, the approach to which would define a continuum limit. In this work we search the parameter space for such a continuum limit.   

An UV limit is obtained by shrinking the lattice spacing $a$ to zero while simultaneously keeping observable quantities fixed in physical units. To give physical meaning to the process of taking an UV limit we must therefore specify the observable we are fixing in physical units. In this work we use fluctuations of three-volume about de Sitter space and the scaling of the spectral dimension as our physical observables.

Using these two independent methods we find the lattice spacing to be strongly dependent on $\kappa_{0}$, but either independent or only very weakly dependent on $\Delta$. Our results suggest one must maximise $\kappa_{0}$ within phase C in order to take a continuum limit. The C-A transition line defines the set of points for which $\kappa_{0}$ is maximised for each $\Delta$ value within phase C. We therefore propose that one must tune to the C-A transition line in order to take a continuum limit. Using the same logic an IRFP may also then exist when $\kappa_{0}$ is minimised along the transition dividing phase C from the bifurcation phase (see Fig. \ref{PDcdtRG}). If this picture is correct a unique RG trajectory would likely follow the path indicated in Fig. \ref{PDcdtRG}. This picture is consistent with the finding of Refs. \cite{Ambjorn:2014gsa,Coumbe:2014noa}, however further study is needed to confirm or refute this proposal. 




\end{section}


\section*{Acknowledgements}

DNC, JGS and JJ wish to acknowledge the support of the grant DEC-2012/06/A/ST2/00389 from the National Science Centre Poland. JA and DNC wish to acknowledge support from the ERC-Advance grant 291092, ``Exploring the Quantum Universe'' (EQU).



\begin{appendices}
\renewcommand{\theequation}{A-\arabic{equation}}
\setcounter{equation}{0}

\section*{Appendix 1}\label{App1}
We derive the proportionality factor in Eq. (\ref{lpl2}) to be
$$
 G = \frac{\sqrt{C_4} s_0^2}{3 \sqrt{6}} \cdot \Gamma \cdot a_{abs}^2 , \nonumber
$$
by analysing quantum fluctuations of the spatial volume about de Sitter space, as first proposed in Refs. \cite{Ambjorn:2007jv,Ambjorn:2008wc}.

We analyse the relation between (dimensionful) continuous and (dimensionless) discrete spatial volume profiles:
\begin{equation}\label{V3Ap}
\langle  V_{3}\left(\tau\right)\rangle= 2 \pi^2 { \cal R}^3 \text{cos}^{3}\left(\frac{\sqrt{g_{\tau\tau}}\ \tau}{ {\cal R}}\right)  =\frac{3}{4}\frac{V_4}{\tilde s_0 V_4^{1/4}}\text{cos}^{3}\left(\frac{ \sqrt{g_{\tau\tau}}\ \tau}{\tilde s_0 V_4^{1/4}}\right) 
\end{equation}
\begin{equation}\label{N3Ap}
\langle N_{3}\left(t\right)\rangle=\frac{3}{4}\frac{N_{(4,1)}}{ s_0 N_{(4,1)}^{1/4}}\text{cos}^{3}\left(\frac{t}{ s_0 N_{(4,1)}^{1/4}}\right) 
\end{equation}
and actions:
\begin{equation}\label{SMSAp}
S_{MS}=\frac{1}{24 \pi G}\int d\tau \sqrt{g_{\tau\tau}} \left(  \frac{ {g^{\tau\tau}} \left( { \partial_\tau V_3(\tau)}  \right)^2}{V_3(\tau)}+ \tilde \mu V_3(\tau)^{1/3}-\tilde \lambda V_3(\tau) \right)  
\end{equation}
\begin{equation}\label{SeffAp}
 S_{eff}= \frac{1}{\Gamma}\sum_t \left(\frac{\Big( N_3(t+1)-N_3(t)\Big)^2}{N_3(t+1)+N_3(t)} + \mu N_3(t)^{1/3} - \lambda N_3(t) \right) ,
\end{equation}
respectively. In the above expressions ${\cal R}$ is the physical radius of the Euclidean de Sitter space (four-sphere) and
$$
V_4 = \int d\tau  \sqrt{g_{\tau\tau}} V_3(\tau) = \frac{8 \pi^2}{3} {\cal R}^4 \quad \Rightarrow \quad \tilde s_0 =  \left (\frac{3}{8 \pi^2}\right)^{1/4} \ .
$$
We also have
$$
\sum_t N_3(t) = N_{(4,1)} \ .
$$
We assume that the CDT universe (inside phase C) is represented by Euclidean de Sitter space with superimposed quantum fluctuations of the scale factor $a(\tau)$, with a spatially homogeneous and isotropic metric
$$
ds^2 = g_{\tau\tau} d\tau^2 + a^2(\tau) d\Omega^3 , \nonumber
$$
where $ d\Omega^3$ is a line element on a $S^3$ sphere.

\subsection*{Method (a)}
Let us assume that consecutive spatial layers of integer time ($t$ and $t+1$)  are separated by a universal lattice distance of constant length:
$$
a_{time}=\sqrt{\tilde\alpha} \cdot a_{abs} \ \ \ (\tilde\alpha = \text{positive const}) .
$$
By construction, a spatial layer in  time $t$ is built of $\frac{1}{2}N_3(t)$ equilateral tetrahedra\footnote{See footnote 1.}, each with equal and constant 3-volume $C_3\cdot a_{abs}^3$, where $C_3 = \frac{\sqrt{2}}{12}$ is the volume of a unit tetrahedron. 
By construction, the total 4-volume of the CDT universe is
\begin{equation}\label{V4Anz}
V_4  = \sum_t  \frac{1}{2} N_3(t) \ C_3 \ a_{abs}^3 \ a_{time}  = \ N_{(4,1)} \ C_4 \ a_{abs}^4,
\end{equation}
where 
\begin{equation}\label{C4a}
C_4 \equiv \frac{\sqrt{ \tilde \alpha} \ C_3}{2} \ .
\end{equation}
Let us assume that\footnote{Note that, as $\tau$ is just a (continuous) time coordinate, the physical proper time is given by $ \sqrt{g_{\tau\tau}} \tau$ and for the physical 3-volume  one should  use  $\sqrt{g_{\tau\tau}} \ V_3(\tau)$ rather than $V_3(\tau) $ .}
\begin{equation}\label{V3Anzatz}
N_3({t}) = \frac{\sqrt{g_{\tau\tau}} \ V_3(\tau)}{C_4 \ a_{abs}^3} \ .
\end{equation}
By applying  ansatz (\ref{V3Anzatz}) and relation (\ref{V4Anz}) to formula (\ref{N3Ap}) for the discrete volume profile one indeed obtains
\begin{equation}\label{VolprofAp}
 \langle V_{3}\left(\tau\right)\rangle =\frac{3}{4}\frac{V_4}{\left(\frac{ \sqrt{g_{\tau\tau}}\ s_0}{C_4^{1/4}}\right) V_4^{1/4}}\text{cos}^{3}\left(\frac{\sqrt{g_{\tau\tau}}\ a_{abs} \ t}{ \left(\frac{\sqrt{g_{\tau\tau}}\  s_0}{C_4^{1/4}}\right) V_4^{1/4}}\right) \ ,
\end{equation}
which compared to the continuous expression  (\ref{V3Ap}) naturally leads to the identifications
\begin{equation}\label{tauAp}
\tau \equiv a_{abs}\ t
\end{equation}
\begin{equation}\label{gttAp}
\frac{\sqrt{g_{\tau\tau}}\  s_0}{C_4^{1/4}} \equiv \tilde s_0 =  \left (\frac{3}{8 \pi^2}\right)^{1/4}.
\end{equation}
Substituting ansatz (\ref{V3Anzatz}) into the  effective action (\ref{SeffAp}) one obtains
$$
 S_{eff}= \frac{1}{\Gamma}\sum_t \frac{\Big( N_3(t+1)-N_3(t)\Big)^2}{N_3(t+1)+N_3(t)} + ...  
= \frac{g_{\tau\tau}}{ \ \Gamma \ C_4 \ a_{abs}^2} \sum_t \Delta \tau \sqrt{g_{\tau\tau}} \frac{\left(\frac{ V_3(\tau+\Delta \tau)-V_3(  \tau )}{\sqrt{g_{\tau\tau}} \Delta \tau} \right)^2}{V_3(\tau+\Delta \tau)+V_3(\tau)} + ...  = 
$$
\begin{equation}\label{Seffa}
= \frac{g_{\tau\tau}}{2 \ \Gamma \ C_4 \ a_{abs}^2}\int d \tau \sqrt{g_{\tau\tau}} \frac{g^{\tau\tau} \left( { \partial_\tau V_3(\tau)}  \right)^2}{V_3(\tau)}+ ... \ ,
\end{equation}
where we used the identifications $\Delta \tau \equiv a_{abs} $ and $ \sum_t \Delta \tau \  \leftrightarrow \ \int d \tau $. Comparing Eqs. (\ref{SMSAp}) and (\ref{Seffa}) implies
\begin{equation}\label{GrelA}
\frac{g_{\tau\tau}}{2 \ \Gamma \ C_4 \ a_{abs}^2} = \frac{1}{24 \pi  G} \ ,
\end{equation}
which combined with  Eq. (\ref{gttAp}) gives the final formula
\begin{equation}\label{Ga}
 G = \frac{\sqrt{C_4} \ s_0^2}{3 \sqrt{6}} \cdot \Gamma \cdot a_{abs}^2\ , 
\end{equation}
where (as defined in Eq. (\ref{C4a}))
$$
C_4 = \frac{\sqrt{2 \tilde \alpha}}{24}  =\text{const}.
$$

 The absolute lattice spacing calculated using  formula (\ref{Ga}) depends on the choice of the dimensionless parameter $\tilde \alpha$ which defines the (temporal) lattice spacing between  neighbouring  spatial layers. In this work we are interested in the relative change in lattice spacing when moving between different points in the CDT parameter space, a result that does not depend on the particular choice of $\tilde \alpha$. However, in order to compare the absolute lattice spacing with the results obtained by using methods (b) and (c) described below, we choose $\tilde \alpha = 5 / 4$, which sets $C_4 = \sqrt{5}/96$ , i.e. equal to the 4-volume of the equilateral unit 4-simplex. 

\subsection*{Method (b)}
We now assume a more realistic picture, where the CDT universe is built of $N_{(4,1)}$ identical (4,1) simplices and  $N_{(3,2)}$ identical (3,2) simplices each with physical volume $C_{41} \cdot a_{abs}^4$ and $C_{32} \cdot a_{abs}^4$, respectively. Therefore, Eq. (\ref{V4Anz}) changes to
\begin{equation}\tag{A-5'}
V_4   = \left( N_{(4,1)} \ C_{41} +  N_{(3,2)} \ C_{32} \right) a_{abs}^4 = \ N_{(4,1)} \ C_4 \ a_{abs}^4 , 
\end{equation}
where
\begin{equation}\tag{A-6'}
C_4 \equiv    C_{41} + \xi  \ C_{32}   \quad , \quad \xi \equiv \frac{N_{(3,2)}}{N_{(4,1)}} \ .
\end{equation}

\begin{figure}[H]
\centering
\includegraphics[width=0.6\linewidth,natwidth=610,natheight=642]{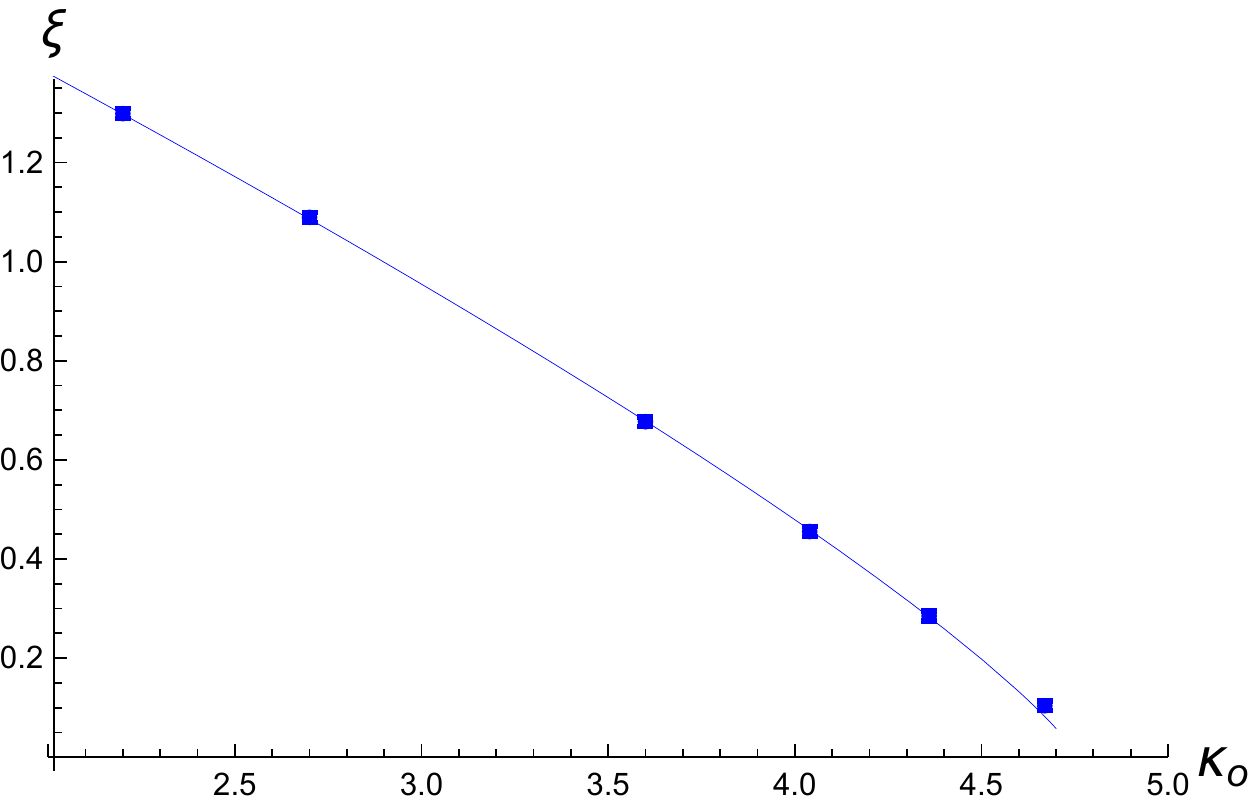}
\caption{\small The dependence  of $\xi \equiv \frac{N_{(3,2)}}{N_{(4,1)}}$ on $\kappa_0$ for fixed $\Delta = 0.6$. $\xi\to 0$ as one approaches the C-A phase transition observed for  $\kappa_0^c \approx 4.7$.  }
\label{ChartXi}
\end{figure}

The rest of the derivation presented in Method (a) remains intact, so one again arrives at formula (\ref{Ga}). Let us now make a simplifying assumption that all 4-simplices are symmetric and
$$
C_{41}=C_{32}=\frac{\sqrt{5}}{96} \ ,
$$ 

\noindent where the numerical value is the volume of a unit equilateral 4-simplex. As a result we obtain

$$
C_4 = \frac{\sqrt{5}}{96} (1+\xi) .
$$
Note, that for  $\xi = 0$ (zero (3,2) simplices) one recovers  the result of Method (a). Actually, as shown in Fig. \ref{ChartXi},  $\xi \to 0$ as one approaches the C-A phase transition, thus  close to the transition the result of Methods (a) and (b) are very similar.


\subsection*{Method (c)}
As in method (b) we assume that the CDT universe is built of both (4,1) and (3,2) simplices but we additionally  take into account that $C_{41}\neq C_{32}$. The volume of a  4-simplex is a function of the asymmetry parameter $\alpha$, defining the ratio  of the length of time-like and space-like links on the lattice 
($a_t^2 = \alpha \cdot a_{s}^2 \equiv \alpha \cdot a_{abs}^2 $), and so one has \cite{Ambjorn:2001cv} \footnote{Here we use the Wick rotated (Euclidean) version of the formulae in \cite{Ambjorn:2001cv}, i.e. the one obtained by an analytical continuation of square roots in the lower half of the complex $\alpha$ plane: $\sqrt{-\alpha}=-i\sqrt \alpha$ . }
\begin{equation}\label{C41C32}
C_{41}=\frac{\sqrt{8\alpha - 3}}{96} \quad , \quad C_{32}=\frac{\sqrt{12\alpha - 7}}{96} \ .
\end{equation}
The asymmetry parameter $\alpha$ is also related to  the bare coupling constants $\kappa_0, \Delta$ and $\kappa_4$ in the bare Regge-Einstein-Hilbert action of CDT (\ref{SRegge})
$$
S_{E}=-\left(\kappa_{0}+6\Delta\right)N_{0}+\kappa_{4}\left(N_{(4,1)}+N_{(3,2)}\right)+\Delta \ N_{(4,1)} 
$$
by the following set of equations \cite{Ambjorn:2001cv}:
\begin{eqnarray}\label{couplingsreal1}
\kappa_0+6\Delta & = & \frac{1}{8  G}\sqrt{4  \alpha-1} \ , \\ 
 \kappa_4+\Delta & = & \frac{\Lambda}{8 \pi G} \frac{\sqrt{8 \alpha-3}}{96} +  \frac{\sqrt 3}{8 \pi G}  \left( \arccos \frac{1}{\sqrt{24 \alpha-8}} - \frac{\pi}{2}  \right) +\\
 &+&\frac{\sqrt{{4  \alpha-1}}}{8 \pi G}  \left(    \frac{3}{2}\arccos\frac{2 \alpha-1}{{6 \alpha-2}}  - \frac{\pi}{2} \right) \ , \nonumber \\\label{couplingsreal3}
 \kappa_4 & = & \frac{\Lambda} {8 \pi G} \frac{\sqrt{12 \alpha-7}}{96} + \frac{\sqrt {3}}{32 \pi G}  \arccos \frac{6 \alpha-5} {6 \alpha-2} + \\
&+ &\frac{ \sqrt{4 \alpha-1}}  {8 \pi G}  \left(  \frac{ 3}{4} \arccos  \frac{4 \alpha-3}{{8 \alpha-4}}  + \frac{3}{2}  \arccos \frac{1}{\sqrt{8 \alpha-4}\sqrt{6 \alpha-2}} -  \pi \right)\  , \nonumber
\end{eqnarray}
where $G$ and $\Lambda$ are the bare Newton's and cosmological constants, respectively. For given values of $\kappa_0, \Delta$ and $\kappa_4$ one can solve Eqs. (\ref{couplingsreal1})-(\ref{couplingsreal3}) for $\alpha$ and then use it to calculate $C_{41}$ and $C_{32}$ according to Eq. (\ref{C41C32}). Summing up, once again we obtain formula (\ref{Ga}) but now we have
$$
C_4=\frac{\sqrt{8\alpha - 3}}{96}+ \xi \frac{\sqrt{12\alpha - 7}}{96},
$$ 
where $\alpha = \alpha(\kappa_0, \Delta, \kappa_4) $. Unfortunately, it turns out that close to the C-A phase transition (for large $\kappa_0$), double valued or complex $\alpha$ solutions  are possible for Eqs.   (\ref{couplingsreal1})-(\ref{couplingsreal3}). As a result method (c) is only valid well inside phase C.

\end{appendices}

\bibliographystyle{unsrt}
\bibliography{Master}

\end{document}